
\input phyzzx
\catcode`\@=11
\font\fourteenmib =cmmib10 scaled\magstep2  \skewchar\fourteenmib='177
\font\twelvemib   =cmmib10 scaled\magstep1   \skewchar\twelvemib='177
\font\tenmib      =cmmib10  \skewchar\tenmib='177
\font\fourteenbsy  =cmbsy10 scaled\magstep2  \skewchar\fourteenbsy='60
\font\twelvebsy    =cmbsy10 scaled\magstep1	  \skewchar\twelvebsy='60
\font\elevenbsy    =cmbsy10 scaled\magstephalf  \skewchar\elevenbsy='60
\font\tenbsy       =cmbsy10    \skewchar\tenbsy='60
\newfam\mibfam
\def\mib{\n@expand\f@m\mibfam}
\let\tmpfourteenf@nts=\fourteenf@nts
\def\fourteenf@nts{\tmpfourteenf@nts %
 \textfont\mibfam=\fourteenmib \scriptfont\mibfam=\tenmib
 \scriptscriptfont\mibfam=\tenmib }
\let\tmptwelvef@nts=\twelvef@nts
\def\twelvef@nts{\tmptwelvef@nts %
 \textfont\mibfam=\twelvemib \scriptfont\mibfam=\tenmib
 \scriptscriptfont\mibfam=\tenmib }
\let\tmptenf@nts=\tenf@nts
\def\tenf@nts{\tmptenf@nts %
    \textfont\mibfam=\tenmib   \scriptfont\mibfam=\tenmib
    \scriptscriptfont\mibfam=\tenmib }
\mathchardef\alpha     ="710B
\mathchardef\beta      ="710C
\mathchardef\gamma     ="710D
\mathchardef\delta     ="710E
\mathchardef\epsilon   ="710F
\mathchardef\zeta      ="7110
\mathchardef\eta       ="7111
\mathchardef\theta     ="7112
\mathchardef\iota      ="7113
\mathchardef\kappa     ="7114
\mathchardef\lambda    ="7115
\mathchardef\mu        ="7116
\mathchardef\nu        ="7117
\mathchardef\xi        ="7118
\mathchardef\pi        ="7119
\mathchardef\rho       ="711A
\mathchardef\sigma     ="711B
\mathchardef\tau       ="711C
\mathchardef\upsilon   ="711D
\mathchardef\phi       ="711E
\mathchardef\chi       ="711F
\mathchardef\psi       ="7120
\mathchardef\omega     ="7121
\mathchardef\varepsilon="7122
\mathchardef\vartheta  ="7123
\mathchardef\varpi     ="7124
\mathchardef\varrho    ="7125
\mathchardef\varsigma  ="7126
\mathchardef\varphi    ="7127

\def\fourteenpoint{\fourteenf@nts \samef@nt \b@gheight=14pt \setstr@t }
\def\twelvepoint{\twelvef@nts \samef@nt \b@gheight=12pt \setstr@t }
\def\tenpoint{\tenf@nts \samef@nt \b@gheight=10pt \setstr@t }
\def\Tenpoint{\tenpoint\twelv@false\spaces@t}
\def\Twelvepoint{\twelvepoint\twelv@true\spaces@t}
\Twelvepoint  
\catcode`\@=12
%
\catcode`\@=11
\def\relaxnext@{\let\next\relax}
\def\hexnumber@#1{\ifcase#1 0\or1\or2\or3\or4\or5\or6\or7\or8\or9\or
 A\or B\or C\or D\or E\or F\fi}
\edef\bffam@{\hexnumber@\bffam}
\mathchardef\boldGamma="0\bffam@00
\mathchardef\boldDelta="0\bffam@01
\mathchardef\boldTheta="0\bffam@02
\mathchardef\boldLambda="0\bffam@03
\mathchardef\boldXi="0\bffam@04
\mathchardef\boldPi="0\bffam@05
\mathchardef\boldSigma="0\bffam@06
\mathchardef\boldUpsilon="0\bffam@07
\mathchardef\boldPhi="0\bffam@08
\mathchardef\boldPsi="0\bffam@09
\mathchardef\boldOmega="0\bffam@0A
\font\tenmsx=msam10
\font\sevenmsx=msam7
\font\fivemsx=msam5
\font\tenmsy=msbm10
\font\sevenmsy=msbm7
\font\fivemsy=msbm5
\newfam\msxfam
\newfam\msyfam
\textfont\msxfam=\tenmsx
\scriptfont\msxfam=\sevenmsx
\scriptscriptfont\msxfam=\fivemsx
\textfont\msyfam=\tenmsy
\scriptfont\msyfam=\sevenmsy
\scriptscriptfont\msyfam=\fivemsy
\font\teneuf=eufm10
\font\seveneuf=eufm7
\font\fiveeuf=eufm5
\newfam\euffam
\textfont\euffam=\teneuf
\scriptfont\euffam=\seveneuf
\scriptscriptfont\euffam=\fiveeuf
\def\frak{\relaxnext@\ifmmode\let\next\frak@\else
 \def\next{\Err@{Use \string\frak\space only in math mode}}\fi\next}
\def\goth{\relaxnext@\ifmmode\let\next\frak@\else
 \def\next{\Err@{Use \string\goth\space only in math mode}}\fi\next}
\def\frak@#1{{\frak@@{#1}}}
\def\frak@@#1{\noaccents@\fam\euffam#1}

%
\edef\msx@{\hexnumber@\msxfam}
\edef\msy@{\hexnumber@\msyfam}
\mathchardef\boxdot="2\msx@00
\mathchardef\boxplus="2\msx@01
\mathchardef\boxtimes="2\msx@02
\mathchardef\square="0\msx@03
\mathchardef\blacksquare="0\msx@04
\mathchardef\centerdot="2\msx@05
\mathchardef\lozenge="0\msx@06
\mathchardef\blacklozenge="0\msx@07
\mathchardef\circlearrowright="3\msx@08
\mathchardef\circlearrowleft="3\msx@09
\mathchardef\leftrightharpoons="3\msx@0B
\mathchardef\boxminus="2\msx@0C
\mathchardef\Vdash="3\msx@0D
\mathchardef\Vvdash="3\msx@0E
\mathchardef\vDash="3\msx@0F
\mathchardef\twoheadrightarrow="3\msx@10
\mathchardef\twoheadleftarrow="3\msx@11
\mathchardef\leftleftarrows="3\msx@12
\mathchardef\rightrightarrows="3\msx@13
\mathchardef\upuparrows="3\msx@14
\mathchardef\downdownarrows="3\msx@15
\mathchardef\upharpoonright="3\msx@16

\mathchardef\downharpoonright="3\msx@17
\mathchardef\upharpoonleft="3\msx@18
\mathchardef\downharpoonleft="3\msx@19
\mathchardef\rightarrowtail="3\msx@1A
\mathchardef\leftarrowtail="3\msx@1B
\mathchardef\leftrightarrows="3\msx@1C
\mathchardef\rightleftarrows="3\msx@1D
\mathchardef\Lsh="3\msx@1E
\mathchardef\Rsh="3\msx@1F
\mathchardef\rightsquigarrow="3\msx@20
\mathchardef\leftrightsquigarrow="3\msx@21
\mathchardef\looparrowleft="3\msx@22
\mathchardef\looparrowright="3\msx@23
\mathchardef\circeq="3\msx@24
\mathchardef\succsim="3\msx@25
\mathchardef\gtrsim="3\msx@26
\mathchardef\gtrapprox="3\msx@27
\mathchardef\multimap="3\msx@28
\mathchardef\therefore="3\msx@29
\mathchardef\because="3\msx@2A
\mathchardef\doteqdot="3\msx@2B

\mathchardef\triangleq="3\msx@2C
\mathchardef\precsim="3\msx@2D
\mathchardef\lesssim="3\msx@2E
\mathchardef\lessapprox="3\msx@2F
\mathchardef\eqslantless="3\msx@30
\mathchardef\eqslantgtr="3\msx@31
\mathchardef\curlyeqprec="3\msx@32
\mathchardef\curlyeqsucc="3\msx@33
\mathchardef\preccurlyeq="3\msx@34
\mathchardef\leqq="3\msx@35
\mathchardef\leqslant="3\msx@36
\mathchardef\lessgtr="3\msx@37
\mathchardef\backprime="0\msx@38
\mathchardef\risingdotseq="3\msx@3A
\mathchardef\fallingdotseq="3\msx@3B
\mathchardef\succcurlyeq="3\msx@3C
\mathchardef\geqq="3\msx@3D
\mathchardef\geqslant="3\msx@3E
\mathchardef\gtrless="3\msx@3F
\mathchardef\sqsubset="3\msx@40
\mathchardef\sqsupset="3\msx@41
\mathchardef\vartriangleright="3\msx@42
\mathchardef\vartriangleleft ="3\msx@43
\mathchardef\trianglerighteq="3\msx@44
\mathchardef\trianglelefteq="3\msx@45
\mathchardef\bigstar="0\msx@46
\mathchardef\between="3\msx@47
\mathchardef\blacktriangledown="0\msx@48
\mathchardef\blacktriangleright="3\msx@49
\mathchardef\blacktriangleleft="3\msx@4A
\mathchardef\vartriangle="0\msx@4D
\mathchardef\blacktriangle="0\msx@4E
\mathchardef\triangledown="0\msx@4F
\mathchardef\eqcirc="3\msx@50
\mathchardef\lesseqgtr="3\msx@51
\mathchardef\gtreqless="3\msx@52
\mathchardef\lesseqqgtr="3\msx@53
\mathchardef\gtreqqless="3\msx@54
\mathchardef\Rrightarrow="3\msx@56
\mathchardef\Lleftarrow="3\msx@57
\mathchardef\veebar="2\msx@59
\mathchardef\barwedge="2\msx@5A
\mathchardef\doublebarwedge="2\msx@5B
\mathchardef\measuredangle="0\msx@5D
\mathchardef\sphericalangle="0\msx@5E
\mathchardef\varpropto="3\msx@5F
\mathchardef\smallsmile="3\msx@60
\mathchardef\smallfrown="3\msx@61
\mathchardef\Subset="3\msx@62
\mathchardef\Supset="3\msx@63
\mathchardef\Cup="2\msx@64

\mathchardef\Cap="2\msx@65

\mathchardef\curlywedge="2\msx@66
\mathchardef\curlyvee="2\msx@67
\mathchardef\leftthreetimes="2\msx@68
\mathchardef\rightthreetimes="2\msx@69
\mathchardef\subseteqq="3\msx@6A
\mathchardef\supseteqq="3\msx@6B
\mathchardef\bumpeq="3\msx@6C
\mathchardef\Bumpeq="3\msx@6D
\mathchardef\lll="3\msx@6E

\mathchardef\ggg="3\msx@6F

\mathchardef\circledS="0\msx@73
\mathchardef\pitchfork="3\msx@74
\mathchardef\dotplus="2\msx@75
\mathchardef\backsim="3\msx@76
\mathchardef\backsimeq="3\msx@77
\mathchardef\complement="0\msx@7B
\mathchardef\intercal="2\msx@7C
\mathchardef\circledcirc="2\msx@7D
\mathchardef\circledast="2\msx@7E
\mathchardef\circleddash="2\msx@7F
\def\ulcorner{\delimiter"4\msx@70\msx@70 }
\def\urcorner{\delimiter"5\msx@71\msx@71 }
\def\llcorner{\delimiter"4\msx@78\msx@78 }
\def\lrcorner{\delimiter"5\msx@79\msx@79 }
\mathchardef\lvertneqq="3\msy@00
\mathchardef\gvertneqq="3\msy@01
\mathchardef\nleq="3\msy@02
\mathchardef\ngeq="3\msy@03
\mathchardef\nless="3\msy@04
\mathchardef\ngtr="3\msy@05
\mathchardef\nprec="3\msy@06
\mathchardef\nsucc="3\msy@07
\mathchardef\lneqq="3\msy@08
\mathchardef\gneqq="3\msy@09
\mathchardef\nleqslant="3\msy@0A
\mathchardef\ngeqslant="3\msy@0B
\mathchardef\lneq="3\msy@0C
\mathchardef\gneq="3\msy@0D
\mathchardef\npreceq="3\msy@0E
\mathchardef\nsucceq="3\msy@0F
\mathchardef\precnsim="3\msy@10
\mathchardef\succnsim="3\msy@11
\mathchardef\lnsim="3\msy@12
\mathchardef\gnsim="3\msy@13
\mathchardef\nleqq="3\msy@14
\mathchardef\ngeqq="3\msy@15
\mathchardef\precneqq="3\msy@16
\mathchardef\succneqq="3\msy@17
\mathchardef\precnapprox="3\msy@18
\mathchardef\succnapprox="3\msy@19
\mathchardef\lnapprox="3\msy@1A
\mathchardef\gnapprox="3\msy@1B
\mathchardef\nsim="3\msy@1C
\mathchardef\ncong="3\msy@1D

\mathchardef\varsubsetneq="3\msy@20
\mathchardef\varsupsetneq="3\msy@21
\mathchardef\nsubseteqq="3\msy@22
\mathchardef\nsupseteqq="3\msy@23
\mathchardef\subsetneqq="3\msy@24
\mathchardef\supsetneqq="3\msy@25
\mathchardef\varsubsetneqq="3\msy@26
\mathchardef\varsupsetneqq="3\msy@27
\mathchardef\subsetneq="3\msy@28
\mathchardef\supsetneq="3\msy@29
\mathchardef\nsubseteq="3\msy@2A
\mathchardef\nsupseteq="3\msy@2B
\mathchardef\nparallel="3\msy@2C
\mathchardef\nmid="3\msy@2D
\mathchardef\nshortmid="3\msy@2E
\mathchardef\nshortparallel="3\msy@2F
\mathchardef\nvdash="3\msy@30
\mathchardef\nVdash="3\msy@31
\mathchardef\nvDash="3\msy@32
\mathchardef\nVDash="3\msy@33
\mathchardef\ntrianglerighteq="3\msy@34
\mathchardef\ntrianglelefteq="3\msy@35
\mathchardef\ntriangleleft="3\msy@36
\mathchardef\ntriangleright="3\msy@37
\mathchardef\nleftarrow="3\msy@38
\mathchardef\nrightarrow="3\msy@39
\mathchardef\nLeftarrow="3\msy@3A
\mathchardef\nRightarrow="3\msy@3B
\mathchardef\nLeftrightarrow="3\msy@3C
\mathchardef\nleftrightarrow="3\msy@3D
\mathchardef\divideontimes="2\msy@3E
\mathchardef\varnothing="0\msy@3F
\mathchardef\nexists="0\msy@40
\mathchardef\mho="0\msy@66
\mathchardef\eth="0\msy@67
\mathchardef\eqsim="3\msy@68
\mathchardef\beth="0\msy@69
\mathchardef\gimel="0\msy@6A
\mathchardef\daleth="0\msy@6B
\mathchardef\lessdot="3\msy@6C
\mathchardef\gtrdot="3\msy@6D
\mathchardef\ltimes="2\msy@6E
\mathchardef\rtimes="2\msy@6F
\mathchardef\shortmid="3\msy@70
\mathchardef\shortparallel="3\msy@71
\mathchardef\smallsetminus="2\msy@72
\mathchardef\thicksim="3\msy@73
\mathchardef\thickapprox="3\msy@74
\mathchardef\approxeq="3\msy@75
\mathchardef\succapprox="3\msy@76
\mathchardef\precapprox="3\msy@77
\mathchardef\curvearrowleft="3\msy@78
\mathchardef\curvearrowright="3\msy@79
\mathchardef\digamma="0\msy@7A
\mathchardef\varkappa="0\msy@7B
\mathchardef\hslash="0\msy@7D
\mathchardef\backepsilon="3\msy@7F
\def\accentfam@{7}
\def\noaccents@{\def\accentfam@{0}}
\catcode`\@=\active

\catcode`\@=11
\paperfootline={\hss\iffrontpage\else\ifp@genum\tenrm
 -- \folio\ --\hss\fi\fi}
\def\titlestyle#1{\par\begingroup \titleparagraphs
 \iftwelv@\fourteenpoint\fourteenbf\else\twelvepoint\twelvebf\fi
 \noindent #1\par\endgroup }
\def\GENITEM#1;#2{\par \hangafter=0 \hangindent=#1
 \Textindent{#2}\ignorespaces}
\def\address#1{\par\kern 5pt\titlestyle{\twelvepoint\sl #1}}
\def\abstract{\par\dimen@=\prevdepth \hrule height\z@ \prevdepth=\dimen@
 \vskip\frontpageskip\centerline{\fourteencp Abstract}\vskip\headskip }
\newif\ifYUKAWA  \YUKAWAtrue
\font\elevenmib   =cmmib10 scaled\magstephalf   \skewchar\elevenmib='177
\def\YUKAWAmark{\hbox{\elevenmib
 Yukawa\hskip0.05cm Institute\hskip0.05cm Kyoto \hfill}}
\def\titlepage{\FRONTPAGE\papers\ifPhysRev\PH@SR@V\fi
 \ifYUKAWA\null\vskip-1.70cm\YUKAWAmark\vskip0.6cm\fi
 \ifp@bblock\p@bblock \else\hrule height\z@ \rel@x \fi }

\def\schapter#1{\par \penalty-300 \vskip\chapterskip
 \spacecheck\chapterminspace
 \chapterreset \titlestyle{\ifcn@@\S\ \chapterlabel.~\fi #1}
 \nobreak\vskip\headskip \penalty 30000
 {\pr@tect\wlog{\string\chapter\space \chapterlabel}} }

\def\ssection#1{\par \ifnum\lastpenalty=30000\else
 \penalty-200\vskip\sectionskip \spacecheck\sectionminspace\fi
 \gl@bal\advance\sectionnumber by 1
 {\pr@tect
 \xdef\sectionlabel{\ifcn@@ \chapterlabel.\fi
 \the\sectionstyle{\the\sectionnumber}}%
 \wlog{\string\section\space \sectionlabel}}%
 \noindent {\S \caps\thinspace\sectionlabel.~~#1}\par
 \nobreak\vskip\headskip \penalty 30000 }


\papers

\def\lkakko{\vbox{\vskip0.065cm\hbox{(}\vskip-0.065cm}}
\def\rkakko{\vbox{\vskip0.065cm\hbox{)}\vskip-0.065cm}}
\def\YUKAWAHALL{\hbox to \hsize
 {\hfil \lkakko\twelvebf YUKAWA HALL\rkakko\hfil}}


\def\Endline{\hfil\break}


\def\addeqno{\ifnum\equanumber<0 \global\advance\equanumber by -1
 \else \global\advance\equanumber by 1\fi}


\mathchardef\Lag="724C
\def\sqr#1#2{{\vcenter{\hrule height.#2pt
 \hbox{\vrule width.#2pt height#1pt \kern#1pt\vrule width.#2pt}
 \hrule height.#2pt}}}

\def\square{{\mathchoice{\sqr84}{\sqr84}{\sqr{5.0}3}{\sqr{3.5}3}}}
\def\pref#1{\rlap.\attach{#1)}}
\def\cref#1{\rlap,\attach{#1)}}
\def\ref#1{\attach{#1)}}



\newdimen\ex@
\ex@.2326ex
\def\boxed#1{\setbox\z@\hbox{$\displaystyle{#1}$}\hbox{\lower.4\ex@
 \hbox{\lower3\ex@\hbox{\lower\dp\z@\hbox{\vbox{\hrule height.4\ex@
 \hbox{\vrule width.4\ex@\hskip3\ex@\vbox{\vskip3\ex@\box\z@\vskip3\ex@}%
 \hskip3\ex@\vrule width.4\ex@}\hrule height.4\ex@}}}}}}
\def\txtboxed#1{\setbox\z@\hbox{{#1}}\hbox{\lower.4\ex@
 \hbox{\lower3\ex@\hbox{\lower\dp\z@\hbox{\vbox{\hrule height.4\ex@
 \hbox{\vrule width.4\ex@\hskip3\ex@\vbox{\vskip3\ex@\box\z@\vskip3\ex@}%
 \hskip3\ex@\vrule width.4\ex@}\hrule height.4\ex@}}}}}}
\newdimen\exx@
\exx@.1ex
\def\thinboxed#1{\setbox\z@\hbox{$\displaystyle{#1}$}\hbox{\lower.4\exx@
 \hbox{\lower3\exx@\hbox{\lower\dp\z@\hbox{\vbox{\hrule height.4\exx@
 \hbox{\vrule width.4\exx@\hskip3\exx@%
 \vbox{\vskip3\ex@\box\z@\vskip3\exx@}%
 \hskip3\exx@\vrule width.4\exx@}\hrule height.4\exx@}}}}}}

\chardef\fontD="1A

\catcode`@=12
%
\pubnum={YITP/K-1089}
\date={October 1994}

\titlepage
\title{Generations of Quarks and Leptons from Noncompact Horizontal
Symmetry}

\author{Kenzo Inoue}
\address{
Yukawa Institute for Theoretical Physics\break
Kyoto University,~Kyoto 606,~Japan}


\vskip 3cm
\noindent
{\bf Abstract}

The three chiral generations of quarks and leptons may be generated
through a spontaneous breakdown of the noncompact horizontal gauge
symmetry $G_{H}$ which governs, together with the standard gauge
symmetry ${SU(3) \times SU(2) \times U(1)}$, the world in a
vectorlike manner.
In a framework of supersymmetric theory, the simplest choice
${G_{H}=SU(1,1)}$ works quite well for this scenario in which quarks,
leptons and Higgses belong to infinite dimensional unitary representation
of $SU(1,1)$.
The relevance of the scenario to the hierarchical structure of their
Yukawa coupling matrices are discussed.

\endpage


\noindent
\S{\bf 1.  ~~Introduction}

One of the remarkable, or even puzzling, facts in the low energy particle
physics is the well-regulated repetition of three generations of quarks
and leptons

$$ q_{i} \equiv {u_i \choose d_i}~,~~ \bar u_{i}~,~~
\bar d_i~, ~~~\ell_{i} \equiv {\nu_i \choose e_i}~,
\bar e_i~, ~~~~~~~~i=1, 2, 3.
\eqno{(1)}
$$

Also mysterious is the well-ordered hierarchical structure of the
coupling matrices ${(y_{u}, ~y_{d}, ~y_{e})}$ of their Yukawa
interactions with Higgs scalars $h$ and $h'$

$$y_u^{ij} ~\bar u_i ~q_{j}~h + y_d^{ij} ~\bar d_i ~q_{j}~h' +
y_e^{ij} ~\bar{e}_{i} ~\ell_{j} ~h'.
\eqno{(2)} $$

Many attempts have been made to find out the basic structure of Nature
lying behind such characteristics of the low energy world\pref{1,2}
Especially the systematic analyses of the Yukawa coupling matrices have
been extensively made\pref{3}
Nevertheless we are still far from the satisfactory understanding of
what is realized in Nature.

Although supersymmetry is now seriously expected to play an important
role in the physics beyond the standard model\cref{4} we are not yet
aware
of its relevance to these problems.
What is worse, in the supersymmetric theory , we must further
understand the reason why the baryon-number and lepton-number violating
Yukawa-type interactions,

$$ \ell \ell \bar e + q \ell \bar d + \bar d \bar d \bar u +
h' h' \bar e ~,
\eqno{(3)} $$

\noindent
are strongly suppressed in the superpotential\pref{5}
Even in the superstring theory\ref{6} at hand, these are only the
criteria for the selection of adequate vacuum from tremendous number of
string vacua.
It does not give any profound understanding of what is realized in
Nature.

One definite approach to these problems, which inquires an
inter-generation structure of quarks and leptons, is to invoke a
symmetry which governs the generation, that is, horizontal
symmetry\pref{2}
Some of the above problems may be understood as a direct consequence of
the symmetry, and the others may be attributed to the spontaneous
breakdown of the symmetry.

It has been for a long time the belief in particle physics that Nature
is as symmetric as possible at the fundamental level.
If the horizontal symmetry has any essential responsibility to
generations, it cannot be irresponsible to the left-right asymmetry of
the low energy physics, which is nothing but the existence of chiral
generation itself.
It will then be a reasonable expectation that Nature is left-right
symmetric (vectorlike)\ref{7} at the fundamental level, and any
asymmetry, even the number of generations, is due to the spontaneous
breakdown of the horizontal symmetry.
The horizontal symmetry, no matter how largely it is broken, will
manifest itself in a clear way at low energies, especially in the
structure of Yukawa coupling matrices under the support of the
nonrenormalization theorem due to supersymmetry.

In this paper we make an attempt to understand generations based on the
horizontal symmetry along the scenario of ``spontaneous generation of
generations".
We work on the supersymmetric gauge theory where gauge group is the
direct product of the horizontal symmetry $G_H$ and the standard gauge
group
$SU(3) \times SU(2) \times U(1)$.
The scenario requires us to introduce a noncompact group as a horizontal
symmetry.
Based on the simplest group
$G_H = SU(1,1)$,
we attempt a model building.
We show how the hierarchical structure of Yukawa coupling matrices
is realized in the scenario.

\vskip 5mm
\noindent
\S{\bf 2. ~~ Spontaneous generation of generations}

Our basic hypothesis is that Nature is left-right symmetric at the
fundamental level, and the gauge group
${G_H} \times {SU(3)} \times {SU(2)} \times {U(1)}$ ~governs the world
in a vectorlike manner\pref{7}
All particles, represented by left-handed chiral superfields,
thus belong to the totally real representation of the gauge group.
We expect that any left-right asymmetries realized at low energies
come through the spontaneous breakdown of $G_H$.

Let $Q$ be the chiral multiplet which belongs to some representation
of $G_H$ and has quantum numbers of quark doublet $q$ under
$SU(3) \times {SU(2)} \times {U(1)}$.
$Q$ contains as subcomponents three generations of
$q_i ~~(i=1, 2, 3)$
as well as extra doublets $q_{extra}$.
We then demand that there exists the chiral multiplet $\bar Q$
which belongs to the conjugate representation of $Q$ under
$G_H \times {SU(3)} \times {SU(2)} \times {U(1)}$.
The scenario ``spontaneous generation of generations" implies that all
components $ \bar q $s in $\bar Q$ and all $q_{extra}$s in
$Q$ acquire huge Dirac mass terms $\bar q$ $q_{extra}$ through the
spontaneous breakdown of $G_H$ and decouple from low energy physics,
retaining $q_1, ~q_2$ and $q_3$ ~massless.
At a glance, this is impossible, because ``vectorlike" implies that
the total number of $q$s is equal to that of $\bar q$s and that
the appearance of massless $ q $s is always accompanied
by the appearance of the same number of massless $\bar q$s as
far as the total number is finite.
This is not a situation we seek for.
The unique loophole which leads to the realization of the scenario
is that $Q$ and $\bar Q$ belong to the infinite dimensional
representation of $G_H$\pref{8}

This is terrible, not only because we have infinite number of
particles (like string theory) but also because the horizontal symmetry
becomes noncompact if we stick to the unitary representation for
the particles.
It is unclear whether we can treat such type of noncompact gauge theory
based on the conventional field theoretical framework or not.
Nevertheless, we dare to proceed further, postponing a lot of things
to future studies.

In order to extract the essential feature of the scenario as transparent
as possible, we work in this paper on the simplest noncompact
horizontal symmetry, $G_H = SU(1,1)$.

The components of any representation of $SU(1.1)$ are labeled by weights,
the eigenvalues of the third component $H_3$ of the $SU(1,1)$
generators $\lbrace H_1, ~H_2, ~H_3 \rbrace$ which form the algebra

$$[H_1, ~H_2] = -iH_3, ~[H_2, H_3] = iH_1, ~[H_3, H_1]=iH_2.
\eqno{(4)}
$$

The unitary representations are infinite dimensional and classified
to two types (positive and negative) depending on the sign of weights.
The positive representation contains components whose weights run from
some real positive number (lowest weight) $\alpha$ to infinity by
one unit,
$\lbrace \alpha, ~\alpha + 1, ~\alpha + 2, \cdot \cdot \cdot \rbrace.
$
The negative representation is the conjugate of the positive
representation, and then the weights of components are
$ \lbrace -\alpha, ~-\alpha-1, ~-\alpha-2, \cdot \cdot \cdot \rbrace
$
with the highest weight $-\alpha$.

Let us first assume that the quark doublets $q$s belong to the positive
representation
$Q_\alpha$ with the lowest weight $\alpha$.
Then $\bar q$s are assigned to the negative representation
$\bar Q_{-\alpha}$ with same $\alpha$:

$$ \eqalign{
Q_\alpha &       = \lbrace q_\alpha, ~q_{\alpha+1}, ~q_{\alpha+2}, \cdot \cdot
                   \cdot \rbrace~~, \cr
\bar Q_{-\alpha}  & = \lbrace \bar q_{-\alpha}, ~\bar q_{-\alpha-1},~
                   \bar q_{-\alpha-2}, \cdot \cdot \cdot \rbrace~~.  \cr
} \eqno{(5)}
$$

At this stage, the choice of sign of weights is a matter of convention.
Once it is fixed by (5), however, the signs for other multiplets are
physically relevant. As we will see later, all quarks and
leptons ($q, ~\bar u, ~\bar d, ~\ell, ~\bar e$) appearing at low energies
must be assigned to positive representations, and their
conjugates ($\bar q, ~u, ~d, ~\bar \ell, ~e$) to negative
representations.
Thus we have following multiplets:

$$ \eqalign{
& Q_\alpha, ~~{\bar U}_\beta, ~~{\bar D}_\gamma, ~~L_\eta,
{}~~{\bar E}_\lambda~~, \cr
& {\bar Q}_{-\alpha}, ~~U_{-\beta}, ~~D_{-\gamma}, ~~\bar L_{-\eta},
{}~~E_{-\lambda}~~, \cr
}
\eqno{(6)}
$$

\noindent
where $\alpha, ~\beta, ~\gamma, ~\eta$ and $\lambda$ are real positive
numbers.

Now we introduce a multiplet $\Psi$ which is responsible to the
spontaneous breakdown of $SU(1,1)$, and couple it to quarks and leptons
in the superpotential by

$$ x_{_Q} ~Q_\alpha ~\bar Q_{-\alpha} ~\Psi + ~x_{_U}
{}~\bar U_
\beta ~U_{-\beta} ~\Psi + x_{_D} ~\bar D_\gamma ~D_{-\gamma} ~\Psi
+ ~x_{_L} ~L_\eta  ~\bar L_{-\eta} ~\Psi +
{}~x_{_E} ~\bar E_\lambda ~E_{-\lambda} ~\Psi~~.
\eqno{(7)}
$$

$\Psi$ is assumed to be singlet under
$SU(3) \times SU(2) \times U(1)$.
As for $SU(1,1)$, the weights of its components are restricted to be
integral. Suppose that the vacuum of the theory breaks $SU(1,1)$ through
the non-vanishing vacuum expectation value ($v. e. v.$)
$\vev{\psi _{-g}} \not= 0$
of some component $\psi _{-g}$ of $\Psi$.
Through the couplings (7), quarks and leptons get masses.
For $Q$ and $\bar Q$, for example, mass term is given by

$$ x_{_Q} ~Q_\alpha ~\bar Q_{-\alpha} \langle \Psi \rangle =
x_{_Q} ~\vev{ \psi _{-g}}  ~\sum_{n=0}^\infty ~C_n^Q
{}~q_{\alpha + n+ g} ~\bar q_{-\alpha -n},
\eqno{(8)}
$$

\noindent
where $C_n^Q$ is the Clebsch-Gordan coefficient.
It should be noticed that the $SU(1,1)$ invariance requires the
additive conservation of the weights.
Eq.(8) implies that the first $g$ components of $Q$ $(q_\alpha,
{}~q_{\alpha+1}, \cdot \cdot \cdot, q_{\alpha+g-1})$
escape from acquiring mass and stay massless.
All components of $\bar Q$ form Dirac mass terms with the remaining
components of $Q$ and become massive.
Thus the choice $g=3$ with $\vev{\psi _{-3}}  \not= 0$
realizes just three generations of massless quark doublets.
In this way, three generations of massless quarks and leptons
($q, ~\bar u, ~\bar d, ~\ell, ~\bar e$) are realized as the staff
members of low energy theory through coupling (7).
We notice that the common sign choice for $Q, ~\bar U, ~\bar D,
{}~L$ and $\bar E$ presented in (6) is essential for this result.
If, for example, $\bar U$ were assigned to negative
representation, $u$ were realized, instead of $\bar u$, in the massless
spectrum.

Up to now, we have not fixed the $SU(1,1)$ representation of
$\Psi$.
If we insist on the unitary nature of the representation, we may assign
it to infinite dimensional representation.
But this is not allowed. The $SU(1,1)$ invariance of the couplings
(7) is realized only when $\Psi$ is assigned to finite dimensional
representation, which is consequently non-unitary.
Thus we have

$$ \Psi = \lbrace
\psi _{-R}, ~\psi _{-R+1}, ~\cdot ~\cdot ~\cdot~,
{}~\psi _0,  ~\cdot ~\cdot ~\cdot~, ~
\psi _{R-1}, ~\psi _R \rbrace~~,
\eqno{(9)}
$$

\noindent
where the highest weight $R$ must be three or more ($R=3,~4,~5, \cdot
\cdot \cdot$) so that $\Psi$ contains the component $\psi _{-3}$.
The remarkable feature we observe from the formula (A8) given in the
Appendix is that the Clebsch-Gordan coefficient $C_n^Q$ in Eq. (8)
behaves as $C_n^Q \sim n^R$ for large $n$, and therefore the masses of
$q$s and $\bar q$s blow up rapidly in $n$.
We thus expect that the infinite number of redundant quarks and
leptons will safely decouple from low energies.

The necessity of the introduction of the non-unitary representation
for $\Psi$ and originally of the introduction of the noncompact gauge
group itself makes it decisive that we cannot work in the conventional
framework of the renormalizable supersymmetric gauge theory.
The canonical kinetic Lagrangian inevitably induces intractable
negative norm states for those belonging to non-unitary representations.
We must work on the theory with non-canonical and consequently
non-renormalizable kinetic Lagrangian.
We expect that the supergravity theory\ref{9}
gives well-defined working
frame for such a theory.
The metric $K_i^j$ in the kinetic Lagrangian of $\Psi$, for example,

$$
K_i^j ~\partial_\gamma ~\psi _i^* ~\partial^\gamma ~\psi _j~~,
\eqno{(10)}
$$

\noindent
is given in terms of the K\"ahler potential $K$ by

$$ K_i^j = {{\partial^2 K} \over {\partial \psi _i^* \partial \psi _j}}~~.
\eqno{(11)}
$$

What is needed is that $K_i^j$ is field dependent and
its $v.~e.~v.$ $\vev{K_i^j}$
is positive definite.
In this context, it is reasonable to expect that $v .~e. ~v.$
of $\psi $ is of order of the Planck mass $M_P$.

The scenario may be phrased in the following way.
Nature has the symmetry $SU(1,1)$ at the fundamental level.
Leptons and quarks belong to infinite dimensional unitary
representations.
There also exist fields $\Psi$ which belong to non-unitary
representations.
The vacuum inevitably breaks $SU(1,1)$ in order to realize
well-defined positive norm states, and leads to non-vanishing
$v.~e.~v.$ for $\Psi$.
In the supersymmetric theory, it is a familiar experience that several
supersymmetric vacua degenerate\pref{4}
If the vacuum with $\vev{\psi _0} \not= 0$
were realized, we would have no chiral generations.
In this
sense, chiral three generations are generated spontaneously
through $\vev{\psi _{-3}} \not= 0$.

Here we emphasize the special role of supersymmetry in the scenario.
If the theory does not have supersymmetry, $\Psi$ is merely a scalar
field belonging to real representation of $SU(1,1)$, and when the
Yukawa coupling $Q \bar Q \Psi$ is allowed by gauge symmetry, the
coupling $Q \bar Q \Psi^\dagger $ is also allowed.
If both terms coexist, the chiral nature of the resulting mass
terms is destroyed and all $q$s become massive.
The supersymmetry forbids the latter coupling due to the chiral nature
of $Q, \bar Q$ and $\Psi$.

\vskip 5mm
\noindent
\S{\bf 3. ~~Higgs multiplets}

Let us now proceed to the Higgs sector.
Supersymmetry requires us to introduce two types of Higgs doublets
$h$ and $h'$ in order to realize Yukawa couplings (2) in the
superpotential\pref{4}
Since all conventional quarks and leptons have positive weights of
$SU(1,1)$,
$h$ and $h'$ must belong to infinite dimensional representations
with negative weights:

$$
\eqalign{
H_{-\rho} & = \lbrace h_{-\rho}, ~h_{-\rho -1}, ~h_{-\rho -2}, \cdot \cdot
              \cdot ~\rbrace~~, \cr
H'_{-\sigma} & = \lbrace ~h'_{-\sigma} ~h'_{-\sigma -1}, ~
                 h'_{\sigma -2}, \cdot
                 \cdot \cdot
                 ~\rbrace~~. \cr
}
\eqno{(12)}
$$

The $SU(1,1)$ invariance of the Yukawa couplings (in the
superpotential)

$$ y_{_U} ~\bar U_{\beta} ~Q_\alpha ~H_{-\rho}
+ y_{_D} ~\bar D_\gamma ~Q_\alpha ~H'_{-\sigma} +
y_{_E} ~\bar E_\lambda ~L_\eta ~h'_{-\sigma} \eqno{(13)}$$

\noindent
restricts possible values of the weights as

$$ \eqalign{
\rho & = \alpha + \beta + \Delta~~ , \cr
\sigma & = \alpha + \gamma + {\Delta}' = \eta + \lambda +
{\Delta}''~~, \cr
}
\eqno{(14)}
$$

\noindent
where $\Delta, {\Delta}'$ and ${\Delta}''$ are non-negative integers
($ 0, 1, 2, ~\cdot ~\cdot ~\cdot $).
The Clebsch-Gordan decomposition of the first term of (13), for example,
takes the form

$$ y_{_U} ~{\bar U}_\beta ~Q_\alpha ~H_{-\rho}
= y_{_U} ~\sum_{i,j=0}^{\infty} ~C_{i,j}^U ~\bar u_{\beta + i}
{}~q_{\alpha+j} ~h_{-\rho + \Delta -i -j}~~.
\eqno{(15)}
$$

The left-right symmetry of the theory requires the existence of the
conjugates

$$
\eqalign{
\bar H_\rho & = \lbrace \bar h_\rho, ~\bar h_{\rho+1},
{}~\bar h_{\rho+2}, \cdot \cdot \cdot \rbrace \cr
\bar H'_\sigma & = \lbrace \bar h'_\sigma, ~\bar h'_{\sigma+1},
{}~\bar h'_{\sigma+2}, \cdot \cdot \cdot \rbrace , \cr
}
\eqno{(16)}
$$

\noindent
which have Yukawa couplings

$$
y_{_U}^* ~U_{-\beta} ~\bar Q_{-\alpha} ~\bar H_\rho
+ y_{_D}^* ~D_{-\gamma} ~\bar Q_{-\alpha} ~{\bar H'}_\sigma
+ y_{_E}^* ~E_{-\lambda} \bar L_\eta ~\bar H'_\sigma~~.
\eqno{(17)}
$$

According to the minimal supersymmetric standard model\cref{4}
we wish to
reproduce just one pair of Higgs doublets $h$ and $h'$ as the
massless states.
For this purpose we introduce a finite dimensional multiplet $\Phi$
with the highest weight $S~~(S=1, 2, 3, \cdot \cdot \cdot),$

$$
\Phi = \lbrace \phi_{-S}, \phi_{-S+1}, ~\cdot \cdot \cdot,
\phi_0, ~\cdot \cdot \cdot, \phi_{S-1}, \phi_S \rbrace~~,
\eqno{(18)}
$$

\noindent
and couple it to Higgs multiplets by

$$
x_{_H} ~H_{-\rho} ~\bar H_\rho ~\Phi + x_{_{H'}} ~H'_{-\sigma}
{}~{\bar H'}_\sigma ~\Phi~~.
\eqno{(19)} $$

The nonvanishing $v.e.v. \vev{\phi_1} \not= 0$ then picks up $h_-\rho$
and $h'_{-\sigma}$ as massless states, making all other
components massive.

For the economical purpose, we might have coupled $\Psi$ of (9) to
$H \bar H$ and $H' \bar H'$ instead of $\Phi$.
In this case, however, the massless states are the first three
components of $\bar H$ and $\bar H'$, which do not have couplings to
massless quarks and leptons.
So we need to introduce at least two multiplets $\Psi$ and $\Phi$
as those which are responsible to the spontaneous breakdown of
$SU(1,1)$.

In this way, all staffs of low energy supersymmetric standard model are
prepared as the chiral massless states. They are

$$
\eqalign{
& q_\alpha, ~q_{\alpha+1}, ~q_{\alpha+2}
{}~~~~~~~~~~\ell_\eta, ~\ell_{\eta+1}, \ell_{\eta+2} \cr
& \bar u_{_\beta}, ~\bar u_{_\beta+1}, ~\bar u_{_\beta+2}
{}~~~~~~~~~~\bar e_\lambda, ~\bar e_{\lambda+1}, ~\bar e_{\lambda+2} \cr
& \bar d_\gamma, ~\bar d_{\gamma+1}, ~\bar d_{\gamma+2}
{}~~~~~~~~~~h_{-\rho}, ~h'_{-\sigma}~~. \cr
}
\eqno{(20)}
$$

The Clebsch-Gordan decomposition (15) and its counterparts for the
other two couplings in (13) completely determine the Yukawa couplings
of these massless states in terms of the Clebsch-Gordan coefficients
$ C_{i,j}^U$ etc.
The detailed structure strongly depends on the values of
$ \Delta, {\Delta}'$ and $ {\Delta}''$ in Eq.(14).
For the $u$-quark coupling

$$
\sum_{i,j=0}^2 \Gamma_u^{ij} ~\bar u_{\beta+i} ~q_{\alpha+j} ~h_{-\rho},
\eqno{(21)}$$

\noindent
we obtain, up to overall normalization,

$$
\eqalign{
& \Gamma_u (\Delta = 0)   \sim \left (\matrix{1 & 0 & 0 \cr
                                      0 & 0 & 0 \cr
                                      0 & 0 & 0 \cr}
                              \right)
\cr
\noalign{\vskip 3mm}
& \Gamma_u(\Delta = 1)  \sim \left (\matrix{0 & -\sqrt \beta & 0 \cr
                                      \sqrt \alpha & 0 & 0 \cr
                                      0 & 0 & 0 \cr}
                               \right)
\cr
\noalign{\vskip 3mm}
& \Gamma_u(\Delta = 2) \sim \left (\matrix{0 & 0 & \sqrt {\beta(2\beta+1)}
                                      \cr
                                      0 & -\sqrt {(2\alpha+1)(2\beta+1)}
                                      & 0 \cr
                                      \sqrt {\alpha(2\alpha+1)} & 0 & 0
                                      \cr}
                                  \right)
\cr
\noalign{\vskip 3mm}
& \Gamma_u(\Delta = 3) \sim \left (\matrix{0 & 0 & 0   \cr
                                           0 & 0 & -\sqrt{2\beta+1} \cr
                                           0 & \sqrt{2\alpha+1} & 0 \cr}
                                   \right)
\cr
\noalign{\vskip 3mm}
& \Gamma_u(\Delta=4)\sim \left (\matrix{0 & 0 & 0 \cr
                                   0 & 0 & 0 \cr
                                   0 & 0 & 1 \cr}
                          \right)
\cr
\noalign{\vskip 3mm}
& \Gamma_u(\Delta \geq 5) \sim 0 ~~.
\cr
}
\eqno{(22)}
$$

We may be tempted to take the third case in (22) $(\Delta = \Delta'=
\Delta'' = 2)$ to realize the observed masses of quarks and leptons
by tuning the values of weights $(\alpha, \beta, \cdot \cdot \cdot)$.
This, however, cannot give the non-trivial Cabibbo-Kobayashi-Maskawa
matrix.
Moreover, we primarily expect that the weights are order one quantity
and are not extremely small nor extremely large.
Thus none of the patterns in(22) looks realistic.
In the next section, we will see that the possible existence of the
$SU(1,1)$ invariant mass terms,
which have been overlooked, gives significant modification.
In this context, the patterns (22) are regarded as the zeroth order
form of the Yukawa couplilng matrix of the model.

\vskip 5mm
\noindent
{\bf \S4.} ~~${\mib SU({\bf 1,1})}$ {\bf invariant mass and generation mixing}

We have constructed a model step by step and have arrived at the
superpotential

$$ \eqalign{W & = x_{_Q} ~Q ~\bar Q ~\Psi + x_{_U} ~\bar U
{}~U \Psi +
x_{_D} ~\bar D ~D ~\Psi + x_{_L} ~L ~\bar L ~\Psi + x_{_E} ~\bar E ~E \Psi \cr
& +x_{_H} ~H ~\bar H ~\Phi + x_{_{H'}} ~H' ~\bar H' \Phi \cr
& +y_{_U} ~\bar U ~Q ~H + y_{_D} ~\bar D ~Q ~H' + y_{_E} ~\bar E ~L ~H'
\cr
& +y_{_U} ~U ~\bar Q ~\bar H + y_{_D} ~D ~\bar Q ~\bar H' +
y_{_E} ~E ~\bar L ~\bar H' \cr
& + \widetilde W (\Psi, \Phi)~~, \cr }
\eqno{(23)}
$$

\noindent
where $\widetilde W$ represents possible self-couplings of $\Psi$ and
$\Phi$.
Without loss of generality, all coupling constants $x$s and $y$s
are taken to be real under the left-right symmetry.

Let us now discuss the possible cubic couplings missing in the
superpotential (23).
The gauge invariance may allow other terms.
The remarkable feature of the model is that the dangerous baryon-number
and lepton-number non-conserving interactions enumerated in (3) are
all forbidden by model construction.
The couplings $LL \bar E, QL \bar D$ and $\bar D \bar D \bar U$
are absent simply because all multiplets have positive weights.
The coupling $H' H' \bar E$ is forbidden by the second constraint
of (14).
Due to the vectorlike extension, the model generally admits new couplings

$$  Q ~Q~D + \bar Q ~\bar Q ~\bar D + Q ~\bar U ~\bar L +
\bar Q ~U ~L \eqno{(24)}$$

\noindent
when the weights satisfy suitable relations.
These terms, however, will be harmless, if exist, because they do not
lead to the cubic couplings among
massless states in (20).
The truly dangerous cubic terms are

$$ \eqalign{
&   (L~H~ +~ \bar L ~\bar H) ~(\Psi ~+ ~\Phi)
\cr
+& (Q ~\bar Q ~+~\bar U ~U + ~\bar D ~D ~+ ~L ~\bar L ~ + ~\bar E ~E
) ~\Phi \cr
+& (H ~\bar H + ~H' ~\bar H')~\Psi~~. \cr}
 \eqno{(25)}
$$

These terms must be absent because they will seriously affect the
massless spectrum of the model.
Although the first term may be forbidden by imposing
$ \vert \eta - \rho \vert \not= integer$,
the second and third terms can not be forbidden by the gauge principle.
This is the most unpleasant aspect of the present
$SU(1,1)$ model.
We may also encounter the difficulty if we further admit the higher
power couplings of $\Psi$ and $\Phi$ to quarks, leptons and Higgses.
For example the coupling
$ Q ~\bar Q ~\Psi^n ~(n \geq 2)$
may tend to reproduce $3n$ generations of $q$s, and the
mixture of different powers of $\Psi$ makes the massless spectrum
quite vague.
At present, we have no good reasoning.
We only expect that the absence of these terms is a manifestation of
the profound feature of more fundamental theory which nontrivially
controls the dynamics of $\Psi$ and $\Phi$.

In addition to the cubic couplings in (23), the $SU(1,1)$
invariance allows following mass terms:

$$
\eqalign{
& M_Q ~Q ~\bar Q + M_U ~\bar U ~U + M_D ~\bar D ~D +
M_L ~L ~\bar L +
M_E ~\bar E ~E \cr
 + & M_H ~H ~\bar H + M_{H'} ~H' ~\bar H'~~.
}
\eqno{(26)}
$$

We expect all these masses are of order $M_P$ because we have no reason
to suppress them.
At a glance, one may suspect that these mass terms completely
upset the scenario based on the couplings (7) and (19), because they
give masses to all quarks, leptons and Higgses.
As a matter of fact, the effect of (26) is to replace the original
massless states (20) by the linear combinations which contain
infinite number of higher-weight states.
As we will see below, this mixing effect which utilizes the seesaw
mechanism\ref{10} gives a significant modification to the Yukawa
coupling matrices\ref{11} (22), which may solve the mystery of Yukawa
coupling hierarchy.

Let us first discuss the Higgs sector $ \lbrace H_{-\rho},
\bar H_\rho \rbrace$~
Their mass terms consists of two parts; one is the
$ SU(1,1)$
invariant mass and the other is the mass due to the coupling to
$\vev{\Phi}$,

$$ \eqalign{
& M_H ~H_{-\rho} ~\bar H_\rho + x_{_H} ~H_{-\rho} ~\bar H_\rho
{}~\vev{\Phi} \cr
=& M_H ~\sum_{n=0}^\infty (-1)^n h_{-\rho -n} ~\bar h _{\rho+n}
+ x_{_H} \vev{\phi_1} \sum_{n=0}^\infty C_n^H ~h_{-\rho-n-1}
{}~\bar h_{\rho +n}
{}~~.
\cr
}
\eqno{(27)}
$$

The important point we recall is that the Clebsch-Gordan
coefficient $C_n^H$ is a monotonically increasing function of $n$
and blows up rapidly for large $n$ by $C_n^H \sim n^S$.
This fact assures, no matter how $M_H$ is large, the existence of
just one massless mode $h$ which consists of a linear combination of
$h_{-\rho-n}$ with rapidly decreasing coefficients,

$$ h = \sum_{n=0}^\infty a_n^{H*} ~h_{-\rho -n}. \eqno{(28)} $$

The coefficient $ a_n^H$ satisfies the recursion equation

$$ M_H ~(-1)^n ~C_n^H + x_{_H} ~\vev{\phi_1} ~C_n^H ~a_{n+1}^H
= 0
\eqno{(29)}
$$

\noindent
which assures the orthogonality of $h$ to massive modes.
This gives

$$ a_n^H = \epsilon _H^n ~a_0^H ~\prod_{r=0}^{n-1}
{}~{{(-1)^r} \over {C_r^H}}~~~,
\eqno{(30)}
$$

\noindent
where

$$ \epsilon _H = - {{M_H} \over {x_{_H} \vev{\phi_1}}}
\eqno{(31)}
$$

\noindent
and $a_0^H$ is fixed by the normalization
$\sum_{n=0}^\infty ~a_n^H ~a_n^{H*} = 1$.

The almost same discussion applies to the other Higgs sector
$ \lbrace H'_{-\sigma}, \bar H'_\sigma \rbrace $,
and reproduces the massless mode $h'$.
For quarks and leptons, the mass terms are

$$ \eqalign{ & M_F ~F ~\bar F + ~x_{_F} ~F ~\bar F \vev{\Psi} \cr
            = & M_F ~\sum_{n=0}^\infty ~(-1)^n ~f_{\alpha_F+n}
{}~\bar f_{-\alpha_F-n}
+x_{_F} ~\vev{\psi _{-3}} ~\sum_{n=0}^\infty ~C_n^F ~f_{\alpha_F+n+3}
{}~\bar f_{-\alpha_F-n}~~, \cr}
 \eqno{(32)}
$$

\noindent
where $F \equiv \lbrace f_{\alpha_F}, f_{\alpha_F + 1}, \cdot \cdot
\cdot \rbrace $
represents $ Q, ~\bar U, ~\bar D, ~L$ and $\bar E$.
Eq.(32) reproduces three massless modes, each of which consists of a
linear combination of the components with weights connected by
three units,

$$ f^{(i)} = \sum_{n=0}^\infty ~a_n^{F(i)*} ~f_{\alpha_F+3n +i}~~,
{}~~~~~i=0,~1,~2.
\eqno{(33)}
$$

The recursion equation similar to (29) gives

$$ a_n^{F(i)} = \epsilon _F^n ~a_0^{F(i)} ~\prod_{r=0}^{n-1}
{}~{{(-1)^{3r+i}} \over {C_{3r+i}^F}} \eqno{(34)} $$

\noindent
with

$$ \epsilon _F \equiv - {{M_F} \over {x_{_F} \vev{\psi _{-3}}}}~~~.
\eqno{(35)}
$$

Thus we have the modified version of the massless modes in one to
one correspondence to the original ones given in (20).
The Yukawa couplings of these modes are obtained by substituting the
inverse relation of Eqs. (28) and (33) to Eq. (15) etc.
For example, for $\bar u ~q ~h$ coupling

$$ \sum_{i,j=0}^2 ~\Gamma_u^{ij} ~\bar u^{(i)} ~q^{(j)} ~h~~,
\eqno{(36)}
$$

\noindent
we have

$$ \Gamma_u^{ij} = y_{_U} ~\sum_{n, n'=0}^\infty
{}~C_{3n+i,~3n'+j}^U ~a_n^{\bar U(i)} ~a_{n'}^{Q(j)}
{}~a_{3(n+n')+i+j-\Delta}^H~~,
\eqno{(37)}
$$

\noindent
where $\Delta$ is a non-negative integer determined by (14),
and $a_{n < 0}^H \equiv 0$
should be understood.

Eq.(37) gives the final formula for the Yukawa coupling matrix
$ \Gamma_u$.
Other Yukawa coupling matrices $ \Gamma_d$
and $ \Gamma_e$ are
obtained by an appropriate replacement of the weights appearing in
$ C_{i,~j}$ and $a_n$.

\vskip 5mm
\noindent
\S {\bf 5. ~~Yukawa coupling hierarchy}

In \S 3, we obtained the zeroth order form of the Yukawa coupling
matrix (22) which was the simplest manifestation of the $SU(1,~1)$
symmetry.
In the last section, we showed how this matrix is modified under
the existence of the $SU(1,~1)$ invariant mass terms (26),
and derived the final formula (37).
In this section, we argue how the formula is relevant to the observed
mass hierarchy of quarks and leptons.

The characteristic feature of the formula (37) is, as  can be
seen from Eqs. (30) and (34), that the mixing coefficients $a_n$s
behave as $a^F_n \sim \epsilon ^n_F$ and $a^H_n \sim \epsilon ^n_H$ with
rapidly
decreasing coefficients.
If we take $\epsilon _F = \epsilon _H = 0$, Eq. (37) reduces to the zeroth
order
form (22).
Therefore, for not so large value of $\epsilon $s ($\epsilon  \lesssim 1$),
$\Gamma ^{ij}_u$ can be safely expanded in terms of the power series of
$\epsilon $s.
Clearly the first nontrivial power of $\epsilon $ in each entry of $\Gamma _u$
depends on the generation indices $i$ and $j$ ($=$ 0, 1, 2).
This means that the coupling matrix $\Gamma ^{ij}_u$ takes a hierarchical
structure in generation space.
In fact, retaining the lowest nonvanishing power term of $\epsilon $ in
Eq. (37), we obtain
$$
\eqalign{
 & \Gamma _u (\Delta  = 0) ~\sim~
\left(
\matrix{
1    & \epsilon     & \epsilon ^2  \cr
\epsilon    & \epsilon ^2  & \epsilon ^3  \cr
\epsilon ^2 & \epsilon ^3  & \epsilon ^4  \cr}\right)
\cr
\noalign{\vskip3mm}
 & \Gamma _u (\Delta  = 1) ~\sim~
\left(
\matrix{
\epsilon ^3 & 1     & \epsilon   \cr
1    & \epsilon   & \epsilon ^2  \cr
\epsilon    & \epsilon ^2  & \epsilon ^3  \cr}\right)
\cr
\noalign{\vskip3mm}
 & \Gamma _u (\Delta  = 2) ~\sim~
\left(
\matrix{
\epsilon ^2 & \epsilon ^3  & 1  \cr
\epsilon ^3 & 1   & \epsilon   \cr
1    & \epsilon   & \epsilon ^2  \cr}\right)
\cr
\noalign{\vskip3mm}
 & \Gamma _u (\Delta  = 3) ~\sim~
\left(
\matrix{
\epsilon  & \epsilon ^2  & \epsilon ^3  \cr
\epsilon ^2  & \epsilon ^3  & 1  \cr
\epsilon ^3  & 1  & \epsilon  \cr}\right)
\cr
\noalign{\vskip3mm}
 & \Gamma _u (\Delta  = 4) ~\sim~
\left(
\matrix{
\epsilon ^4 & \epsilon      & \epsilon ^2  \cr
\epsilon    & \epsilon ^2  & \epsilon ^3  \cr
\epsilon ^2   & \epsilon ^3  & 1  \cr}\right)~~,
\cr
}
\eqno{(38)}
$$
where we have set $\epsilon _H = \epsilon _Q = \epsilon _{\bar U} \equiv
\epsilon $ and omitted
the numerical coefficients.

The observed mass spectrum of quarks and leptons suggests that the
first case in (38) is realized for all Yukawa couplings, that is,
$\Delta =\Delta '=\Delta ''=0$ in Eq.(14).
Then the masses of the 1st, 2nd and 3rd generation quarks and leptons
are $0(\epsilon ^4)$, $0(\epsilon ^2)$ and $0(1)$.
The Cabibbo-Kobayashi-Maskawa matrix takes the form
$$
U_{CKM} ~\sim~
\left(
\matrix{
1 & \epsilon  & \epsilon ^2 \cr
\epsilon  & 1 & \epsilon  \cr
\epsilon ^2 & \epsilon  & 1  \cr
}\right)~~.
\eqno{(39)}
$$
These results are certainly qualitatively reasonable when we recall
the numerical factors related to the Clebsch-Gordan coefficients.

The fact that the power $\epsilon ^n$ appears in Eq.(37) accompanied by
rapidly decreasing coefficient further strengthens the hierarchical
structure of $\Gamma _u$ due to $\epsilon ^n$ itself.
Let us discuss the case $\Delta =0$ more precisely retaining numerical
factors.
In this case, the terms with $n>0$ and/or $n' > 0$ in (37) always
give the higher order corrections, and the leading contribution is
given by the first term with $n = n'= 0$;
$$
\Gamma ^{ij}_u ~\sim~ y_{_U} ~C^U_{i,j} ~a^H_{i+j}~~.
\eqno{(40)}
$$
The Clebsch-Gordan coefficients are read off from the formulas (A8)
and (A9) given in the Appendix:
$$
\eqalignno{
C^U_{i,~j} &
= N \sqrt{{{(i+j)!~\Gamma (2\alpha +j)~\Gamma (2\beta +i)}\over%
{i!~j!~\Gamma (2\alpha +2\beta +i+j)}}} & (41) \cr
\noalign{\vskip6mm}
C^H_n &
= N'(-1)^n \sqrt{{{n!~(n + 1)!}\over%
{\Gamma (2\rho  + n)~\Gamma (2\rho  + n + 1)}}} & \cr
      &
\times \sum^{S-1}_{r=0}~{{\Gamma (2\rho  + n + 1 + r)}\over%
{(S-r)!~(S-r-1)!~r!~(r+1)!~(r+1+n-S)!}} & (42) \cr
}
$$
where $\rho = \alpha +\beta $, and $N$ and $N'$ are normalization constants.
The integer $S$, which is the highest weight of the multiplet $\Phi $,
is restricted to positive integer.
The growing feature of $C^H_n$ with respect to $n$ becomes much
radical for larger value of $S$, but even in the smallest case
$S=1$, it is still sizable and realizes a remarkable hierarchy in
the numerical coefficients in $\Gamma ^{ij}_u$ :
$$
\Gamma ^{ij}_u ~\sim~ y_{_U} \epsilon _H^{i+j}~
{{\Gamma (2\alpha  + 2\beta )}\over{\Gamma (2\alpha  + 2\beta  + i + j)}}~
\sqrt{{{\Gamma (2\alpha  + j)~\Gamma (2\beta  + i)}\over%
{i!~j!~\Gamma (2\alpha )~\Gamma (2\beta )}}}~~,
\eqno{(43)}
$$
where we have fixed the normalization constants $N$ and $N'$ such
that $C^U_{0,~0} = C^H_0 = 1$.
For example if we take $\alpha  = \beta  = 1/2$, we have
$$
\Gamma ^{ij}_u ~\sim~ y_{_U}~{\epsilon ^{i+j}_H \over (i + j + 1)!}~~.
\eqno{(44)}
$$
The eigenvalues of this matrix are
$$
\eqalign{
& y_{_U}~(1 + 0(\epsilon ^2_H)~)~~, \cr
& y_{_U}~(-{1 \over 12}~\epsilon ^2_H + 0(\epsilon ^4_H)~)~~, \cr
& y_{_U}~({1 \over 720}~\epsilon ^4_H + 0(\epsilon ^6_H)~)~~. \cr
}
\eqno{(45)}
$$
This shows that not so small value of $\epsilon _H$, for example
$\epsilon _H ~\sim~1/3$, is able to reproduce the observed large mass
hierarchy of quarks and leptons.

\vskip5mm
\noindent
\S {\bf 6. ~~Discussions}

In this paper we have made an attempt at the model building based on
the noncompact horizontal gauge group $SU(1,~1)$ following the scenario
of ``spontaneous generation of generations".
This scenario may give new insight on the origin of the chiral
generations of quarks and leptons and their hierarchical Yukawa couplings.
Although it is too premature to decide the viability of the scenario
from the very limitted analysis presented here, we expect that the
gross feature of the obtained results is the reflection of the general
feature of the scenario.
Further studies are much desired both from phenomenological and
theoretical points of view.

The most important phenomenological problem is whether the Yukawa
coupling matrix (37) really reproduces the observed masses of quarks
and leptons and the Cabibbo-Kobayashi-Maskawa matrix $U_{CKM}$ under
the reasonable choice of the weights.
This requires a comprehensive analysis retaining full freedom of
the parameters of the model including the relative phase of $\VEV{\Psi }$ and
$\VEV{\Phi }$ indispensable for the $CP$ violating phase in $U_{CKM}$.
To be precise, we must further discuss the mass term of Higgs multiplets
$h$ and $h'$.
The minimal supersymmetric standard model requires the mass term
$\mu hh'$ in the superpotential.
Such a mass term is, however, protected to vanish by $SU(1,~1)$ even
after it is spontaneously broken.
The unique remedy will be to extend the model so that the extra light
$SU(3) \times SU(2) \times SU(1)$ singlets $s$ survive at low energies
and couple to $h$ and $h'$ in the superpotential by $shh'$.

The theoretical situation is much less satisfactory.
First of all, it is unclear whether we can treat, in a manner adopted
here, the theory of supergravity with noncompact gauge symmetry
containing infinite dimensional unitary multiplets as well as
finite dimensional non-unitary ones.
If such a theory exists, the consistency may require some constraints
on the detailed structure of the theory, especially
on the possible values of the weights.
Furthermore we simply assumed the $v.e.v.$s of $\Psi $ and $\Phi $
following our will.
In principle, they must be determined through the stationality condition
of the bosonic potential, which demands the knowledge on the K\"ahler
potential $K$ as well as the superpotential $W$.
The requirement that these $v.e.v.$s do not break the local
supersymmetry and at the same time realize the positive definite
metric for all particle states will impose severe constraint on the
structure of $K$ and $W$.
The vacuum structure of the theory must be clarified.

In our analysis we implicitly assumed the minimal form of the
K\"ahler potential\ref{9} for quarks, leptons and Higgses.
In general, $\Psi $ and $\Phi $ may be allowed to couple to them freely in
the form like
$Q^{\dagger} f(\Psi ,~\Psi ^{\dagger},~\Phi ,~\Phi ^{\dagger})Q$ as far as they
do not disturb the positivity of the metric.
In this case the Yukawa coupling matrix (37) receives the modification
due to the wave function renormalization of $Q$, $\bar U$ and $H$.
So we need to know the principle which determines the form of
$f(\Psi ,~\Psi ^{\dagger},~\Phi ,~\Phi ^{\dagger})$ in order to derive
the fully reliable results.

The characterization of the model further requires the clarification
of the gauge anomalies\pref{8}
Since we are working on the vectorlike theory, we might not worry about
the gauge anomaly.
However, when $SU(1,~1)$ is spontaneously broken, theory becomes chiral
through the coupling of $\Psi $ and $\Phi $ to matters in the superpotential.
The model has been constructed by hand so that the resulting chiral
fields form the anomaly free sets.
But who knew this?
The consistency of the theory seems to restrict the structure of
the superpotential beyond the gauge invariance at the classical level.

We expect that the future attempts push some of these problems and
open the way to the deeper understanding of the origin of generations
of quarks and leptons.

\vskip9mm
\noindent
{\bf Acknowledgements}

This work is partly supported by the Grant-in-Aid for Scientific
Research from the Ministry of Education, Science and Culture (05640344).

\endpage

\noindent
{\bf Appendix}

In this appendix we give some formulas for the $SU(1,~1)$ invariants
needed in the text.

The $SU(1,~1)$ invariant bi-linears are
$$
\eqalignno{
Q^{\dagger}_\alpha  Q_\alpha   & \equiv \sum^\infty _{n=0}~q^*_{\alpha
+n}~q_{\alpha +n}~~, & (A1) \cr
\noalign{\vskip5mm}
\bar{Q}^{\dagger}_{-\alpha } \bar{Q}_{-\alpha }  & \equiv \sum^\infty
_{n=0}~\bar{q}^*_{-\alpha -n}~\bar{q}_{-\alpha -n}~~, & (A2) \cr
\noalign{\vskip5mm}
Q_\alpha  \bar{Q}_{-\alpha } & \equiv \sum^\infty _{n=0}(-1)^n~ q_{\alpha
+n}~\bar{q}_{-\alpha -n}~~, & (A3) \cr
\noalign{\vskip5mm}
\Psi ^{\dagger} \Psi   & \equiv \sum^R_{n=-R}(-1)^n~ \psi ^*_n~ \psi _n~~, &
(A4) \cr
\noalign{\vskip5mm}
\Psi ~ \Psi '  & \equiv \sum^R_{n=-R}(-1)^n~ \psi _n~ \psi '_{-n}~~, & (A5) \cr
}
$$
where $\Psi $ and $\Psi '$ are assumed to have the highest weight $R$.

For the cubic invariants
$$
\eqalignno{
Q_\alpha  ~\bar{Q}_{-\alpha } ~\Psi  & \equiv \sum^\infty _{i,j=0}~A_{i,~j}
{}~q_{\alpha +i}~\bar{q}_{-\alpha -j}~\Psi _{-i+j}~~, & (A6) \cr
\noalign{\vskip5mm}
\bar{U}_{\beta } ~Q_\alpha  ~H_{-\rho } & \equiv \sum^\infty _{i,j=0}~B_{i,~j}
{}~\bar{u}_{\beta +i} ~q_{\alpha +j}~h_{-\rho +\Delta -i-j}~~, & (A7) \cr
}
$$
where $\Delta \equiv \rho -\alpha -\beta $ is restricted to non-negative
integers,
the Clebsch-Gordan coefficients $A_{i,~j}$ and $B_{i,~j}$ are given by
$$
\eqalign{
A_{i,~j} &
= N_A (-1)^j~\sqrt{{{(i-j+R)!~i!~j!~(-i+j+R)!}\over%
{\Gamma (2\alpha +i)~\Gamma (2\alpha +j)}}} \cr
         &
\times \sum^{i-j+R}_{r=0}~{{\Gamma (2\alpha +j+r)}\over{%
(R-r)! ~(i-j+R-r)! ~r! ~(j-i+R)! ~(r+j-R)!}} \cr
}
\eqno{(A8)}
$$
and
$$
\eqalign{
B_{i,~j} &
= N_B (-1)^{i+j}~\sqrt{{{ i! ~j!~\Gamma (2\beta +i) ~\Gamma (2\alpha +j)}\over%
{(i+j-\Delta )! ~\Gamma (2\rho +i+j-\Delta )}}} \cr
         &
\times \sum^{\Delta }_{r=0} (-1)^r~{{(i+j-\Delta )!}\over{%
(i-r)! ~(j+r-\Delta )! ~r! ~(\Delta -r)! ~\Gamma (2\beta +r)~ \Gamma (2\alpha
-r+\Delta )}} \cr
}
\eqno{(A9)}
$$
where $N_A$ and $N_B$ are normalization constants independent of $i$
and $j$.

\endpage

\noindent
{\bf References}

\item{1)}
F. Wilczek and A. Zee, Phys. Lett. {\bf 70B} (1977), 418. \Endline
H. Fritzsch, Phys. Lett. {\bf 70B} (1977), 436. \Endline
A. De Rujura, H. Georgi and S.L. Glashow, Ann. of Phys. {\bf 109} (1977), 258.
\Endline
S. Pakvasa and H. Sugawara, Phys. Lett. {\bf 73B} (1978), 61. \Endline
R. Barbieri, R. Gatto and F. Strocchi, Phys. Lett. {\bf 74B} (1978), 344.
\Endline
H. Harari, H. Haut and J. Weyers, Phys. Lett. {\bf 78B} (1978), 459. \Endline
H. Georgi and C. Jarlskog, Phys. Lett. {\bf 89B} (1979), 297. \Endline
J. Harvey, P. Ramond and D. Reiss, Phys. Lett. {\bf 92B} (1980), 309. \Endline
Y. Koide, Phys. Rev. {\bf D28} (1983), 252.

\item{2)}
T. Maehara and T. Yanagida, Prog. Theor. Phys. {\bf 61} (1979), 1434. \Endline
F. Wilczek and A. Zee, Phys. Rev. Lett. {\bf 42} (1979), 421.

\item{3)}
S. Dimopoulos, L.J. Hall and S. Raby, Phys. Rev. Lett. {\bf 68} (1992), 1984;
Phys. Rev. {\bf D45} (1992), 4195. \Endline
H. Arason, D.J. Castano, E.J. Piard and P. Ramond, Phys. Rev. {\bf D47}
(1993), 232. \Endline
M. Bando, K. Izawa and T. Takahashi, Prog. Theor. Phys. {\bf 92} (1994),
143.

\item{4)}
S. Dimopoulos and H. Georgi, Nucl. Phys. {\bf B193} (1981), 150. \Endline
N. Sakai, Z. Phys. {\bf C11} (1982), 153. \Endline
K. Inoue, A. Kakuto, H. Komatsu and S. Takeshita, Prog. Theor. Phys.
{\bf 67} (1982), 1889; {\bf 68} (1982), 927. \Endline
L.E. Ibanez and G.G. Ross, Phys. Lett. {\bf 110B} (1982), 215. \Endline
L. Alvarez-Gaume, J. Polchinski and M.B. Wise, Nucl. Phys. {\bf B221}
(1983), 495. \Endline
J. Ellis, J.S. Hagelin, D.V. Nanopoulos and K. Tamvakis, Phys. Lett.
{\bf 125B} (1983), 275. \Endline
H.P. Nills, Phys. Reports {\bf 110} (1984), 1. \Endline
U. Amaldi, W. de Boer and H. Furstenau, Phys. Lett. {\bf B260} (1991),
447. \Endline
J. Ellis, S. Kelley and D.V. Nanopoulos, Phys. Lett. {\bf B260} (1991),
131. \Endline
P. Langacker and M. Luo, Phys. Rev. {\bf D44} (1991), 817.

\item{5)}
S. Weinberg, Phsy. Rev. {\bf D26} (1982), 287. \Endline
N. Sakai and T. Yanagida, Nucl. Phys. {\bf B197} (1982), 533.

\item{6)}
M. Green and J. Schwarz, Phys. Lett. {\bf 149B} (1984), 117. \Endline
D. Gross, J. Harvey, E. Martinec and R. Rohm, Phys. Rev. Lett. {\bf 54}
(1985), 502. \Endline
P. Candelas, G. Horowitz, A. Strominger and E. Witten, Nucl. Phys.
{\bf B 258} (1985), 46.

\item{7)}
F.A. Wilczek, A. Zee, R.L. Kingsley and S.B. Treiman, Phys. Rev.
{\bf D12} (1975), 2768. \Endline
A. De R\'ujula, H. Georgi and S.L. Glashow, Phys. Rev. {\bf D12} (1975),
3589. \Endline
H. Fritzsch, M. Gell-Mann and P. Minkowski, Phys. Lett. {\bf 59B} (1975),
256. \Endline
K. Inoue, A. Kakuto and Y. Nakano, Prog. Theor. Phys. {\bf 58} (1977), 630.
\Endline
M. Yoshimura, Prog. Theor. Phys. {\bf 58} (1977), 972.

\item{8)}
S.A. Frolov and A.A. Slavnov, Nucl. Phys. {\bf B411} (1994), 647.

\item{9)}
D.Z. Freedman, P. van Nieuwenhuizen and S. Ferrara, Phys. Rev. {\bf D13}
(1976), 3214. \Endline
D.Z. Freedman and P. van Nieuwenhuizen, Phys. Rev. {\bf D14} (1976), 912.
\Endline
S. Deser and B. Zumino, Phys. Lett. {\bf 62B} (1976), 335. \Endline
E. Cremmer, B. Julia, J. Scherk, S. Ferrara, L.Girardello and P. van
Nieuwenhuizen, Nucl. Phys. {\bf B147} (1979), 105. \Endline
E. Cremmer, S. Ferrara, L. Girardello and A. van Proeyen, Nucl. Phys.
{\bf B212} (1983), 413. \Endline
T. Kugo and S. Uehara, Nucl. Phys. {\bf B222} (1983), 125.

\item{10)}
T. Yanagida, in Proceedings of the Workshop on the Unified Theories
and the Baryon Number in the Universe, ed. O. Sawada and A. Sugamoto
(KEK, Tsukuba, 1978). \Endline
M. Gell-Mann, P. Ramond and R. Slansky, in Supergravity, ed. P. van
Nieuwenhuizen and D.Z. Freedman (North Holland, Amsterdam, 1979).

\item{11)}
D. Chang and R.N. Mohapatra, Phys. Rev. Lett. {\bf 58} (1987), 1600. \Endline
S. Rajpoot, Phys. Lett. {\bf B191} (1987), 122. \Endline
A. Davidson and K.C. Wali, Phys. Rev. Lett. {\bf 59} (1987), 393. \Endline
K.S. Babu, J.C. Pati and H. Stremnitzer, Phys. Lett. {\bf B256} (1991), 206.

\bye